\documentstyle[12pt]{article}
\begin{document}


\begin{titlepage}

\null

\begin{flushright}
 KOBE-TH-94-01 \\
 March 1994
\end{flushright}

\vspace{1cm}
\begin{center}
 {\Large\bf
 Space-time Supersymmetry in Asymmetric \par}
\vspace{3mm}
 {\Large\bf
 Orbifold Models
 \par}
\vspace{2.5cm}
\baselineskip=7mm
 {\large
  Toshihiro Sasada
  \par}
\vspace{5mm}
{\sl
  Department of Physics, Kobe University\\
  Rokkodai, Nada, Kobe 657, Japan
  \par}

\vspace{3cm}
 {\large\bf Abstract}
\end{center}
\par

We study the condition for the appearance of space-time supercharges in
twisted
sectors of asymmetric orbifold models.
We present a list of the asymmetric $Z_N$-orbifold models which satisfy a
simple condition necessary for the appearance of space-time supercharges in
twisted sectors.
We investigate whether or not such asymmetric orbifold models possess $N=1$
space-time supersymmetry and obtain a new class of $N=1$ asymmetric
orbifold
models.
It is pointed out that the result of space-time supersymmetry does not
depend
on any choice of a shift if the order of the left-moving degrees of freedom
is
preserved.

\end{titlepage}


\baselineskip=7mm


Orbifold compactification \cite{DHVW1,DHVW2} is one of the efficient
methods to
construct phenomenologically realistic string models and systematic
construction of orbifold models has been performed by many groups
\cite{GROUPS}.
Since the heterotic string \cite{GHMR} has left-right asymmetric nature,
asymmetric orbifolds \cite{NSV1,NSV2} are natural extension of symmetric
orbifolds and are expected to supply phenomenologically realistic string
models.
In the construction of four-dimensional string models, the preservation of
$N=1$ space-time supersymmetry seems to be the most reliable condition to
select the phenomenologically viable string models.
For symmetric orbifold models, the condition for the preservation of $N=1$
space-time supersymmetry is given by the simple condition which is written
in
terms of the eigenvalues of the automorphism matrices as discussed in ref.
\cite{DHVW2}.
However, this condition is not applicable to asymmetric orbifold models in
general since there might appear gravitino states in some twisted sectors
of
asymmetric orbifolds\footnote{For symmetric orbifolds, gravitino states do
not
appear from twisted sectors.}.
These gravitino states will correspond to the space-time supercharges from
twisted sectors and as a result the space-time supersymmetry in such
asymmetric
orbifold models will be ``enhanced'' \cite{Symmetries,Supercurrents}.
In this paper, we shall study the condition for the occurrence of the
space-time supersymmetry ``enhancement'' and investigate whether or not
the
asymmetric orbifold models which satisfy the necessary condition for the
space-time supersymmetry ``enhancement'' possess $N=1$ space-time
supersymmetry\footnote{An example of the asymmetric orbifold models which
possesses the ``enhanced'' $N=1$ space-time supersymmetry is discussed in
ref.
\cite{Supercurrents}.}.

In order to investigate the condition for the preservation of $N=1$
space-time
supersymmetry of the fermionic string models, it may be convenient to use
the
bosonic string map \cite{CENST,LLS,LSWrep} and investigate the
Ka\v{c}-Moody
algebras in the right-moving degrees of freedom of the corresponding
bosonic
string models\footnote{In this paper we use only the ``weak'' version
\cite{LLS,LSWrep} of the bosonic string map.}.
In the case of four-dimensional string models, the bosonic string map works
as
follows:
The light-cone $SO(2)$ Ka\v{c}-Moody algebra generated by the NSR fermions
in
the fermionic string theory is replaced by an $SO(10) \times E_8$
Ka\v{c}-Moody
algebra in the bosonic string theory.
This is done in such a way that the $SO(2)$ Ka\v{c}-Moody characters of the
fermionic string theory are mapped to the $SO(10) \times E_8$ Ka\v{c}-Moody
characters of the bosonic string theory preserving the modular
transformation
properties of the Ka\v{c}-Moody characters.
Under this map, the existence of a gravitino state (i.e. an $N=1$
space-time
supercharge) in the fermionic string theory corresponds to the existence of
a
set of operators of the left-right conformal weight (0,1) transforming as a
spinor of the $SO(10)$ in the bosonic string theory.
Then it can be shown that, in the bosonic string theory, such a set of
operators of the conformal weight (0,1) should extend the $SO(10)$
Ka\v{c}-Moody algebra to the $E_6$ Ka\v{c}-Moody algebra \cite{LSWlett}.

Under the bosonic string map, the condition discussed in ref. \cite{DHVW2}
for
the preservation of $N=1$ space-time supersymmetry of symmetric orbifold
models
becomes the condition that there exist the states of the conformal weight
(0,1)
in the untwisted sector of the bosonic symmetric orbifold models and that
the
operators corresponding to them should extend the $SO(10)$ Ka\v{c}-Moody
algebra to the $E_6$ Ka\v{c}-Moody algebra.
The reason for this is that there is no state of the conformal weight (0,1)
in
the twisted sectors of the bosonic symmetric orbifold models since the
left-
and right-conformal weights of the ground states of any twisted sector are
both
positive and equal in the case of the bosonic symmetric orbifold models.
On the other hand, in the case of asymmetric orbifold models there is the
possibility for the existence of the states of the conformal weight (0,1)
in
some twisted sectors of the bosonic asymmetric orbifold models.
The existence of such states implies a symmetry between the untwisted and
twisted sectors and the operators corresponding to the (0,1) states will
enlarge the Ka\v{c}-Moody algebra of the bosonic asymmetric orbifold models
\cite{CHDGM,Twist,String,Symmetries,Supercurrents,Fermion}.
If these operators of the conformal weight (0,1) in some twisted sectors
enlarge the Ka\v{c}-Moody algebra containing the $SO(10)$ Ka\v{c}-Moody
algebra
of the bosonic asymmetric orbifold models, then the space-time
supersymmetry of
the asymmetric orbifold model will be ``enhanced''.
Therefore, the necessary condition for the space-time supersymmetry
``enhancement'' in asymmetric orbifold models is the existence of the
states of
the conformal weight (0,1) in some twisted sectors of the corresponding
bosonic
asymmetric orbifold models.
In the following, we will consider the asymmetric $Z_N$-orbifold models
which
satisfy the necessary condition for the space-time supersymmetry
``enhancement'' and investigate whether or not such asymmetric orbifold
models
possess $N=1$ space-time supersymmetry.

In the construction of an asymmetric orbifold model, we start with a
toroidal
compactification of the $E_8\times E_8$ heterotic string theories which is
specified by a $(22+6)$-dimensional even self-dual lattice
$\Gamma^{16,0}\oplus\Gamma^{6,6}$ \cite{NSW}, where $\Gamma^{16,0}$ is a
root
lattice of $E_8\times E_8$.
The left- and right-moving momentum $(p_L^I, p_L^i, p_R^i)$ $(I=1, \ldots
,16,
i=1, \ldots ,6)$ lies on the lattice $\Gamma^{16,0}\oplus\Gamma^{6,6}$.
Let $g$ be a group element which generates a cyclic group $Z_N$.
The $g$ is defined to act on the left- and right-moving string coordinate
$(X_L^I, X_L^i, X_R^i)$ by
\begin{equation}
g: (X_L^I, X_L^i, X_R^i) \rightarrow (X_L^I + 2\pi v_L^I, U_L^{ij} X_L^j,
U_R^{ij} X_R^j),
\label{tr}
\end{equation}
where $U_L$ and $U_R$ are rotation matrices which satisfy $U_L^N=U_R^N=1$
and
$v_L^I$ is a constant vector which satisfy $Nv_L^I\in \Gamma^{16,0}$.
The $Z_N$-transformation must be an automorphism of the lattice
$\Gamma^{6,6}$,
i.e.,
\begin{equation}
(U_L^{ij} p_L^j, U_R^{ij} p_R^j) \in \Gamma^{6,6} \ \  {\rm for \ all} \
(p_L^i,p_R^i) \in \Gamma^{6,6}.
\end{equation}
The action of the operator $g$ on the right-moving fermions is given by the
$U_R$ rotation.
We denote the eigenvalues of $U_L$ and $U_R$ by $ \{ e^{i 2\pi
\zeta_L^a},e^{-i
2\pi \zeta_L^a} ; a=1,2,3 \} $ and $ \{ e^{i 2\pi \zeta_R^a},e^{-i 2\pi
\zeta_R^a} ; a=1,2,3 \} $, respectively.
The necessary condition for one-loop modular invariance is for $N$ odd
\begin{equation}
N \frac{1}{2} \sum_{a=1}^3 \zeta_L^a (1 - \zeta_L^a) + N \frac{1}{2}
(v_L^I)^2
= 0 \ \  \bmod 1;
\label{mod}
\end{equation}
for $N$ even, in addition to the condition (\ref{mod}) we get the following
conditions:
\begin{equation}
N \sum_{a=1}^3 \zeta_R^a = 0 \ \  \bmod 2,
\end{equation}
\begin{equation}
  p_L^i (U_L^{\frac{N}{2}})^{ij} p_L^j - p_R^i (U_R^{\frac{N}{2}})^{ij}
p_R^j =
0 \ \  \bmod 2
\end{equation}
for all $(p_L^i,p_R^i) \in \Gamma^{6,6}$ \cite{NSV1}.
These are called the left-right level matching conditions and it has been
proved that these are also sufficient conditions for one-loop modular
invariance \cite{V,NSV2}.

Let $N_L$ be the order of the left-moving degrees of freedom, i.e.
$U_L^{N_L} =
1$ and $N_Lv_L^I\in \Gamma^{16,0}$.
In the $g^{\ell}$-twisted sector where $\ell$ is the multiples of $N_L$,
the
left-moving conformal weight of the ground states of the corresponding
bosonic
asymmetric orbifold model is zero since $U_L^{N_L} = 1$ and $N_Lv_L^I\in
\Gamma^{16,0}$.
Thus, there is the possibility for the existence of the states of the
conformal
weight (0,1) in such $g^{\ell}$-twisted sectors of the corresponding
bosonic
asymmetric orbifold model.
Therefore, in the case of the asymmetric $Z_N$-orbifold models we obtain a
simple condition $N_L \neq N$ which is necessary for the space-time
supersymmetry ``enhancement'', where $N_L$ is the order of the left-moving
degrees of freedom.
It should be noted that for our purpose it is sufficient to consider the
``non-degenerate'' element (i.e. $\zeta_R^a \neq 0  $ $(a = 1, 2, 3)$) of
the
right-moving automorphism\footnote{The relation between the
``non-degenerate''
elements of the right-moving automorphisms and space-time chirality is
discussed in ref. \cite{SW1} in the context of the covariant lattice
formulation of four-dimensional strings.}.
The reason for this is that in the case of the ``degenerate'' element (i.e.
$\zeta_R^a = 0$ for some $a$) of the automorphism the operators of the
conformal weight (0,1) in the untwisted sector of the bosonic asymmetric
orbifold model will extend the $SO(10)$ Ka\v{c}-Moody algebra to the
Ka\v{c}-Moody algebra containing the $SO(12)$ Ka\v{c}-Moody algebra and
that
such an algebra containing the $SO(12)$ Ka\v{c}-Moody algebra cannot be
enlarged to the $E_6$ Ka\v{c}-Moody algebra by the operators of the
conformal
weight (0,1) in the twisted sectors of the bosonic asymmetric orbifold
model.
We also note that the result of the space-time supersymmetry
``enhancement''
does not depend on any choice of the shift $v_L^I$ as long as the order of
the
left-moving degrees of freedom $N_L$ is preserved since in the
$g^{\ell}$-twisted sector where $\ell$ is the multiples of $N_L$ the
physical
states of the asymmetric orbifold model are determined irrespective of the
choice of the shift $v_L^I$.

We will now discuss the asymmetric $Z_N$-orbifold models with inner
automorphisms of the momentum lattices associated with simply-laced Lie
algebras.
We take the lattice $\Gamma^{6,6}$ of an asymmetric $Z_N$-orbifold model to
be
of the form:
\begin{equation}
\Gamma^{6,6} = \{ (p_L^i,p_R^i) \vert  p^i_L, p^i_R\in \Lambda^{\ast},
p_L^i-p_R^i \in \Lambda  \},
\label{lat}
\end{equation}
where $\Lambda$ is a 6-dimensional lattice and $\Lambda^{\ast}$ is the dual
lattice of $\Lambda$.
It turns out that $\Gamma^{6,6}$ is Lorentzian even self-dual if $\Lambda$
is
even integral.
In the following, we will take $\Lambda$ in eq. (\ref{lat}) to be the
products
of root lattices of the simply-laced Lie algebras with the squared length
of
the root vectors normalized to two \cite{EN}.
The left- and right-rotation matrices of the $Z_N$-transformation in eq.
(\ref{tr}) are taken from the Weyl group elements of the root lattices of
the
simply-laced Lie algebras.
Then it is easy to see that such a $Z_N$-transformation is an automorphism
of
the lattice $\Gamma^{6,6}$ in eq. (\ref{lat}).
For our purpose, it is sufficient to investigate the ``non-degenerate''
elements (i.e. $\zeta_R^a \neq 0  $ $(a = 1, 2, 3)$) for the right-moving
automorphisms.
On the other hand, for the left-moving automorphisms there is no reason for
restricting our consideration to the ``non-degenerate'' elements (i.e.
$\zeta_L^a \neq 0  $ $(a = 1, 2, 3)$) of the automorphisms and we will
investigate all Weyl group elements including the ``degenerate'' elements
(i.e.
$\zeta_L^a = 0$ for some $a$) of the automorphisms in the left-moving
degrees
of freedom\footnote{Massless states of the four-dimensional string models
arising from the covariant lattices which correspond to the
``non-degenerate''
Weyl group elements in both left- and right-moving degrees of freedom are
investigated in ref. \cite{SW2}.}.
In our consideration the shift $v_L^I$ is chosen as
\begin{equation}
v_L^I = (\zeta_L^a, 0^5; 0^8).
\label{std}
\end{equation}
The choice (\ref{std}) of the shift $v_L^I$ simplifies the condition
(\ref{mod}) for one-loop modular invariance:
For $N$ odd, the condition (\ref{mod}) is always satisfied; for $N$ even,
the
condition (\ref{mod}) reduces to the following simple condition:
\begin{equation}
N \sum_{a=1}^3 \zeta_L^a = 0 \ \  \bmod 2.
\end{equation}

The unified description of the Weyl groups of all simple Lie algebras have
been
discussed in refs. \cite{C,HM,SW1,LSWrep,B}.
Here, we will briefly summarize the results.
Let $w$ be an element of a Weyl group $W$.
Any $w\in W$ has the following expression \cite{C}:
\begin{equation}
w = w_{\alpha _1} \ldots w_{\alpha _k} w_{\alpha _{k+1}} \ldots w_{\alpha
_{k+h}},
\label{dec}
\end{equation}
where $\alpha _1, \ldots, \alpha _{k+h}$ are linearly independent roots and
$\{
\alpha _1, \ldots, \alpha _k \}$, $\{ \alpha _{k+1}, \ldots, \alpha _{k+h}
\}$
are each the set of mutually orthogonal roots.
Corresponding to the decomposition (\ref{dec}), we shall define a
Dynkin-like
graph which is called the Carter diagram.
If $w\in W$ has a decomposition with a graph, any Weyl group element which
is
conjugate to $w$ also has the decomposition with the same graph.
Classification of the graphs associated with the decomposition (\ref{dec})
is
discussed for any simple Lie algebras.
Although the correspondence between the conjugacy classes and the graphs is
not
always one-to-one, the exceptions are fully discussed in ref. \cite{C}.

Although the left- and right-rotations of the $Z_N$-transformations in eq.
(\ref{tr}) can be defined for arbitrary elements of the Weyl groups, the
conditions for modular invariance put restrictions on the left- and
right-rotations of the $Z_N$-transformations.
All the models we have to consider are shown in table 1.
The root lattices $\Lambda$ associated with the momentum lattices
$\Gamma^{6,6}$ of the asymmetric orbifold models are given in the first
column
of table 1.
The left- and right-moving Carter diagrams $C_L$ and $C_R$ of the
automorphisms
of the momentum lattices $\Gamma^{6,6}$ of the asymmetric orbifold models
are
given in the second and the third columns of table 1, respectively.
The gauge groups $G$ of the asymmetric orbifold models are calculated in
the
same way as in refs. \cite{String,Symmetries} by rewriting equivalently the
automorphisms of the lattices $\Gamma^{6,6}$ into shifts in the lattices
\cite{DHVW2,NSV1,S} and the results are given in the fourth column of table
1.
Using the bosonic string map, we investigate the space-time supersymmetry
``enhancement'' of the asymmetric orbifold models.
The number of the space-time supercharges from the untwisted sector of the
asymmetric orbifold models is given in the fifth column of table 1.
The total number of the space-time supercharges from the untwisted and
twisted
sectors of the asymmetric orbifold models is given in the sixth column of
table
1.
We can easily check that, for all the asymmetric orbifold models with $N_L
\neq
N$ investigated in this paper, there always exist the states with the
conformal
weight (0,1) in some twisted sectors of the corresponding bosonic
asymmetric
orbifold models.
These operators of the conformal weight (0,1) in the twisted sectors often
enlarge the Ka\v{c}-Moody algebra containing the $SO(10)$ Ka\v{c}-Moody
algebra
and the space-time supersymmetry  ``enhancement'' occurs in such asymmetric
orbifold models.
Accordingly, we obtain in table 1 a new class of asymmetric orbifold models
with the ``enhanced'' $N=1$ space-time supersymmetry.
On the other hand, in some of the asymmetric orbifold models none of the
operators of the conformal weight (0,1) in the twisted sectors enlarges the
Ka\v{c}-Moody algebra containing the $SO(10)$ Ka\v{c}-Moody algebra and no
space-time supersymmetry ``enhancement'' is observed in such asymmetric
orbifold models.
Accordingly, we also obtain in table 1 a new class of $N=1$ asymmetric
orbifold
models without the space-time supersymmetry ``enhancement'' which possess
the
operators of the conformal weight (0,1) in the twisted sectors of the
corresponding bosonic asymmetric orbifold models.

In this paper we have studied the condition for the space-time
supersymmetry
``enhancement'' in asymmetric orbifold models.
We have presented in table 1 a list of the asymmetric $Z_N$-orbifold models
with inner automorphisms of the momentum lattices satisfying the simple
condition $N_L \neq N$ which is necessary for the space-time supersymmetry
``enhancement'', where $N_L$ is the order of the left-moving degrees of
freedom.
In this list of the asymmetric orbifold models, we have obtained a new
class of
$N=1$ asymmetric orbifold models with or without the space-time
supersymmetry
``enhancement''.
It seems that whether or not the asymmetric orbifold models which satisfy
the
necessary condition for the space-time supersymmetry ``enhancement''
possess
$N=1$ space-time supersymmetry depends on the details of the momentum
lattices
and the automorphisms of the lattices.
Although we have only considered the choice of the shift $v_L^I$ in eq.
(\ref{std}), the results in table 1 of space-time supersymmetry of the
asymmetric $Z_N$-orbifold models are unchanged for the other choice of the
shift $v_L^I$ as long as the order of the left-moving degrees of freedom
$N_L$
is preserved.
For example, it is easy to check that the previously discussed $E_8\times
E_8$
asymmetric orbifold models with the space-time supersymmetry
``enhancement''
\cite{Symmetries,Supercurrents} are all understood by considering the
$v_L^I =
0$ embedding (i.e. no embedding in the gauge degrees of freedom) of the
asymmetric orbifold models discussed in this paper.
(The orbifold models in ref. \cite{Symmetries} are to be regarded only as
illustrative of the features of asymmetric orbifolds and not of a direct
phenomenological relevance.)
It appears that the choice of the shifts in eq. (\ref{std}) and the other
choice of the shifts which satisfy the condition (\ref{mod}) with the order
of
the left-moving degrees of freedom preserved are respectively the most
natural
generalization of the ``standard'' and ``non-standard'' embeddings of
symmetric
orbifolds.
Classification of the shifts $v_L^I$ of asymmetric $Z_N$-orbifolds
consistent
with the condition (\ref{mod}) for one-loop modular invariance will be
carried
out in exactly the same way as has been done in refs. \cite{KKKOOT} for
symmetric $Z_N$-orbifolds.
It would be of great interest to study the asymmetric orbifold models with
such
shifts $v_L^I$.
We hope to get new phenomenologically interesting string models along this
line.

\bigskip

The author would like to thank Dr. M. Sakamoto for reading the manuscript
and
useful comments.



\newpage



\newpage

\pagestyle{empty}


\begin{table}

\caption{
List of asymmetric $Z_N$-orbifold models.
$\Lambda$ denote the root lattices associated with the momentum lattices.
$C_L$ and $C_R$ denote the left- and right-moving Carter diagrams of the
automorphisms of the momentum lattices, respectively, where semicolons
separate
the direct products of the lattices.
$G$ denote the gauge groups of the asymmetric orbifold models where the
hidden
$E_8$ gauge groups are omitted.
$M$ denotes the number of the space-time supercharges from the untwisted
sectors of the asymmetric orbifold models.
$N$ denotes the total number of the space-time supercharges from the
untwisted
and twisted sectors of the asymmetric orbifold models.
}

\vspace{3 mm}

\centering

\begin{tabular}{|c|c|c|c|c|c|}
\hline
$ \Lambda $&$ C_L $&$ C_R $&$ G $&$ M $&$ N $\\
\hline
$ A_2\times A_2\times A_2 $&$ (\phi ;\phi ;\phi ) $&$ (A_2 ;A_2 ;A_2 ) $&$
E_8\times SU(3)^3 $&$ 1 $&$ 2 $\\
$ A_4\times A_2 $&$ (\phi ;\phi ) $&$ (A_4 ;A_2 ) $&$ E_8\times SU(5)\times
SU(3) $&$ 0 $&$ 4 $\\
$ A_4\times A_2 $&$ (\phi ;A_2 ) $&$ (A_4 ;A_2 ) $&$ SO(14)\times
SU(5)\times
U(1)^3 $&$ 0 $&$ 0 $\\
$ A_4\times A_2 $&$ (A_2 ;\phi ) $&$ (A_4 ;A_2 ) $&$ SO(14)\times
SU(3)^2\times
U(1)^3 $&$ 0 $&$ 0 $\\
$ A_4\times A_2 $&$ (A_2 ;A_2 ) $&$ (A_4 ;A_2 ) $&$ E_7\times SU(3)\times
U(1)^5 $&$ 0 $&$ 0 $\\
$ A_4\times A_2 $&$ (A_4 ;\phi ) $&$ (A_4 ;A_2 ) $&$ SO(12)\times
SU(3)\times
U(1)^6 $&$ 0 $&$ 0 $\\
$ A_6 $&$ (\phi ) $&$ (A_6 ) $&$ E_8\times SU(7) $&$ 1 $&$ 4 $\\
$ A_6 $&$ (A_2 ) $&$ (A_6 ) $&$ E_8\times SU(7) $&$ 1 $&$ 4 $\\
$ A_6 $&$ (A_2^2 ) $&$ (A_6 ) $&$ E_8\times SU(7) $&$ 1 $&$ 4 $\\
$ A_6 $&$ (A_4 ) $&$ (A_6 ) $&$ E_8\times SU(7) $&$ 1 $&$ 4 $\\
$ D_4\times A_2 $&$ (\phi ;\phi ) $&$ (D_2^2 ;A_2 ) $&$ E_8\times
SO(8)\times
SU(3) $&$ 0 $&$ 4 $\\
$ D_4\times A_2 $&$ (\phi ;A_2 ) $&$ (D_2^2 ;A_2 ) $&$ SO(14)\times
SO(8)\times
U(1)^3 $&$ 0 $&$ 0 $\\
$ D_4\times A_2 $&$ (\phi ;\phi ) $&$ (D_4(a_1) ;A_2 ) $&$ E_8\times
SO(8)\times SU(3) $&$ 0 $&$ 4 $\\
$ D_4\times A_2 $&$ (\phi ;A_2 ) $&$ (D_4(a_1) ;A_2 ) $&$ SO(14)\times
SO(8)\times U(1)^3 $&$ 0 $&$ 0 $\\
$ D_4\times A_2 $&$ (\phi ;\phi ) $&$ (D_4 ;A_2 ) $&$ E_8\times SO(8)\times
SU(3) $&$ 1 $&$ 4 $\\
$ D_4\times A_2 $&$ (\phi ;A_2 ) $&$ (D_4 ;A_2 ) $&$ SO(14)\times
SO(8)\times
U(1)^3 $&$ 1 $&$ 2 $\\
$ D_4\times A_2 $&$ (D_2^2 ;\phi ) $&$ (D_2^2 ;A_2 ) $&$ E_7\times
SU(3)\times
SU(2)^5 $&$ 0 $&$ 2 $\\
$ D_4\times A_2 $&$ (D_2^2 ;\phi ) $&$ (D_4(a_1) ;A_2 ) $&$ E_7\times
SU(3)\times SU(2)^5 $&$ 0 $&$ 2 $\\
$ D_4\times A_2 $&$ (D_4(a_1) ;\phi ) $&$ (D_2^2 ;A_2 ) $&$ E_7\times
SU(3)\times SU(2)^5 $&$ 0 $&$ 2 $\\
$ D_4\times A_2 $&$ (D_2^2 ;A_2 ) $&$ (D_4(a_1) ;A_2 ) $&$ SO(10)\times
SU(2)^6\times U(1)^3 $&$ 0 $&$ 0 $\\
$ D_4\times A_2 $&$ (A_2 ;\phi ) $&$ (D_2^2 ;A_2 ) $&$ SO(14)\times
SU(3)\times
SU(2)^3\times U(1)^2 $&$ 0 $&$ 0 $\\
$ D_4\times A_2 $&$ (A_2 ;A_2 ) $&$ (D_2^2 ;A_2 ) $&$ E_7\times
SU(2)^3\times
U(1)^4 $&$ 0 $&$ 0 $\\
$ D_4\times A_2 $&$ (D_2^2 ;\phi ) $&$ (D_4 ;A_2 ) $&$ E_7\times
SU(3)\times
SU(2)^5 $&$ 1 $&$ 2 $\\
$ D_4\times A_2 $&$ (D_4(a_1) ;\phi ) $&$ (D_4(a_1) ;A_2 ) $&$ E_7\times
SU(3)\times SU(2)\times U(1)^4 $&$ 0 $&$ 0 $\\
$ D_4\times A_2 $&$ (A_2 ;\phi ) $&$ (D_4(a_1) ;A_2 ) $&$ SO(14)\times
SU(3)\times SU(2)^3\times U(1)^2 $&$ 0 $&$ 0 $\\
$ D_4\times A_2 $&$ (A_2 ;A_2 ) $&$ (D_4(a_1) ;A_2 ) $&$ E_7\times
SU(2)^3\times U(1)^4 $&$ 0 $&$ 0 $\\
$ D_4\times A_2 $&$ (D_4(a_1) ;\phi ) $&$ (D_4 ;A_2 ) $&$ E_7\times
SU(3)\times
SU(2)^5 $&$ 1 $&$ 2 $\\
$ D_4\times A_2 $&$ (D_4 ;\phi ) $&$ (D_4(a_1) ;A_2 ) $&$ SO(12)\times
SU(3)\times U(1)^6 $&$ 0 $&$ 0 $\\
$ D_4\times A_2 $&$ (D_4 ;A_2 ) $&$ (D_4(a_1) ;A_2 ) $&$ E_6\times U(1)^8
$&$ 0
$&$ 0 $\\
$ D_4\times A_2 $&$ (A_2 ;\phi ) $&$ (D_4 ;A_2 ) $&$ SO(14)\times
SU(3)\times
SU(2)^3\times U(1)^2 $&$ 1 $&$ 2 $\\
$ D_4\times A_2 $&$ (A_2 ;A_2 ) $&$ (D_4 ;A_2 ) $&$ E_7\times SU(2)^3\times
U(1)^4 $&$ 1 $&$ 2 $\\
\hline
\end{tabular}

\end{table}


\clearpage

\setcounter{table}{0}


\begin{table}

\caption{
(continued)
}

\vspace{3 mm}

\centering

\begin{tabular}{|c|c|c|c|c|c|}
\hline
$ \Lambda $&$ C_L $&$ C_R $&$ G $&$ M $&$ N $\\
\hline
$ D_6 $&$ (\phi ) $&$ (D_4 (a_1 ) D_2 ) $&$ E_8\times SO(12) $&$ 1 $&$ 4
$\\
$ D_6 $&$ (\phi ) $&$ (D_6 (a_1 ) ) $&$ E_8\times SO(12) $&$ 1 $&$ 4 $\\
$ D_6 $&$ (D_2^2 ) $&$ (D_4 (a_1 ) D_2 ) $&$ E_7\times SO(8)\times SU(2)^3
$&$
1 $&$ 2 $\\
$ D_6 $&$ (D_2^2 ) $&$ (D_6 (a_1 ) ) $&$ E_7\times SO(8)\times SU(2)^3 $&$
1
$&$ 2 $\\
$ D_6 $&$ (A_1^2 D_2 ) $&$ (D_4 (a_1 ) D_2 ) $&$ E_7\times SO(8)\times
SU(2)^3
$&$ 1 $&$ 2 $\\
$ D_6 $&$ (A_1^2 D_2 ) $&$ (D_6 (a_1 ) ) $&$ E_7\times SO(8)\times SU(2)^3
$&$
1 $&$ 2 $\\
$ D_6 $&$ (D_4 (a_1 ) ) $&$ (D_6 (a_1 ) ) $&$ E_7\times SU(4)\times
SU(2)\times
U(1)^3 $&$ 1 $&$ 1 $\\
$ D_6 $&$ (A_2 ) $&$ (D_4 (a_1 ) D_2 ) $&$ E_8\times SO(12) $&$ 1 $&$ 4 $\\
$ D_6 $&$ (D_4 ) $&$ (D_4 (a_1 ) D_2 ) $&$ E_7\times SO(8)\times SU(2)^3
$&$ 1
$&$ 2 $\\
$ D_6 $&$ (A_2^2 ) $&$ (D_4 (a_1 ) D_2 ) $&$ E_8\times SO(12) $&$ 1 $&$ 4
$\\
$ D_6 $&$ (D_4 (a_1 ) D_2 ) $&$ (D_6 (a_1 ) ) $&$ E_6\times SO(8)\times
SU(2)^2\times U(1)^2 $&$ 1 $&$ 1 $\\
$ D_6 $&$ (A_4 ) $&$ (D_4 (a_1 ) D_2 ) $&$ E_8\times SO(12) $&$ 1 $&$ 4 $\\
$ D_6 $&$ (A_2 ) $&$ (D_6 (a_1 ) ) $&$ E_8\times SO(12) $&$ 1 $&$ 4 $\\
$ D_6 $&$ (D_4 ) $&$ (D_6 (a_1 ) ) $&$ E_7\times SO(8)\times SU(2)^3 $&$ 1
$&$
2 $\\
$ D_6 $&$ (A_2^2 ) $&$ (D_6 (a_1 ) ) $&$ E_8\times SO(12) $&$ 1 $&$ 4 $\\
$ D_6 $&$ (A_4 ) $&$ (D_6 (a_1 ) ) $&$ E_8\times SO(12) $&$ 1 $&$ 4 $\\
$ E_6 $&$ (\phi ) $&$ (A_2^3 ) $&$ E_8\times E_6 $&$ 1 $&$ 4 $\\
$ E_6 $&$ (\phi ) $&$ (A_5 A_1 ) $&$ E_8\times E_6 $&$ 1 $&$ 4 $\\
$ E_6 $&$ (\phi ) $&$ (E_6 ) $&$ E_8\times E_6 $&$ 1 $&$ 4 $\\
$ E_6 $&$ (\phi ) $&$ (E_6 (a_1 ) ) $&$ E_8\times E_6 $&$ 0 $&$ 4 $\\
$ E_6 $&$ (\phi ) $&$ (E_6 (a_2 ) ) $&$ E_8\times E_6 $&$ 1 $&$ 4 $\\
$ E_6 $&$ (A_2 ) $&$ (A_5 A_1 ) $&$ SO(14)\times SU(6)\times U(1)^2 $&$ 1
$&$ 1
$\\
$ E_6 $&$ (A_2 ) $&$ (E_6 ) $&$ SO(14)\times SU(6)\times U(1)^2 $&$ 1 $&$ 1
$\\
$ E_6 $&$ (A_2 ) $&$ (E_6 (a_1 ) ) $&$ SO(14)\times SU(6)\times U(1)^2 $&$
0
$&$ 1 $\\
$ E_6 $&$ (A_2 ) $&$ (E_6 (a_2 ) ) $&$ SO(14)\times SU(6)\times U(1)^2 $&$
1
$&$ 2 $\\
$ E_6 $&$ (A_1^4 ) $&$ (A_2^3 ) $&$ E_8\times E_6 $&$ 1 $&$ 4 $\\
$ E_6 $&$ (A_1^4 ) $&$ (A_5 A_1 ) $&$ E_7\times SU(6)\times SU(2)^2 $&$ 1
$&$ 2
$\\
$ E_6 $&$ (A_1^4 ) $&$ (E_6 ) $&$ E_7\times SU(6)\times SU(2)^2 $&$ 1 $&$ 2
$\\
$ E_6 $&$ (A_1^4 ) $&$ (E_6 (a_1 ) ) $&$ E_8\times E_6 $&$ 0 $&$ 4 $\\
$ E_6 $&$ (A_1^4 ) $&$ (E_6 (a_2 ) ) $&$ E_7\times SU(6)\times SU(2)^2 $&$
1
$&$ 2 $\\
\hline
\end{tabular}

\end{table}


\clearpage

\setcounter{table}{0}


\begin{table}

\caption{
(continued)
}

\vspace{3 mm}

\centering

\begin{tabular}{|c|c|c|c|c|c|}
\hline
$ \Lambda $&$ C_L $&$ C_R $&$ G $&$ M $&$ N $\\
\hline
$ E_6 $&$ (A_2^2 ) $&$ (A_5 A_1 ) $&$ E_7\times SO(8)\times U(1)^3 $&$ 1
$&$ 1
$\\
$ E_6 $&$ (A_2^2 ) $&$ (E_6 ) $&$ E_7\times SO(8)\times U(1)^3 $&$ 1 $&$ 1
$\\
$ E_6 $&$ (A_2^2 ) $&$ (E_6 (a_1 ) ) $&$ E_7\times SO(8)\times U(1)^3 $&$ 0
$&$
1 $\\
$ E_6 $&$ (A_2^2 ) $&$ (E_6 (a_2 ) ) $&$ E_7\times SO(8)\times U(1)^3 $&$ 1
$&$
2 $\\
$ E_6 $&$ (A_4 ) $&$ (A_2^3 ) $&$ E_8\times E_6 $&$ 1 $&$ 4 $\\
$ E_6 $&$ (A_4 ) $&$ (A_5 A_1 ) $&$ E_8\times E_6 $&$ 1 $&$ 4 $\\
$ E_6 $&$ (A_4 ) $&$ (E_6 ) $&$ E_8\times E_6 $&$ 1 $&$ 4 $\\
$ E_6 $&$ (A_4 ) $&$ (E_6 (a_1 ) ) $&$ E_8\times E_6 $&$ 0 $&$ 4 $\\
$ E_6 $&$ (A_4 ) $&$ (E_6 (a_2 ) ) $&$ E_8\times E_6 $&$ 1 $&$ 4 $\\
$ E_6 $&$ (D_4 ) $&$ (E_6 ) $&$ SO(12)\times SU(3)^2\times U(1)^4 $&$ 1 $&$
1
$\\
$ E_6 $&$ (D_4 ) $&$ (E_6 (a_1 ) ) $&$ SO(14)\times SU(6)\times U(1)^2 $&$
0
$&$ 1 $\\
$ E_6 $&$ (D_4 (a_1 ) ) $&$ (A_2^3 ) $&$ E_8\times E_6 $&$ 1 $&$ 4 $\\
$ E_6 $&$ (D_4 (a_1 ) ) $&$ (A_5 A_1 ) $&$ E_7\times SU(6)\times SU(2)^2
$&$ 1
$&$ 2 $\\
$ E_6 $&$ (D_4 (a_1 ) ) $&$ (E_6 ) $&$ E_7\times SU(3)^2\times SU(2)\times
U(1)^2 $&$ 1 $&$ 2 $\\
$ E_6 $&$ (D_4 (a_1 ) ) $&$ (E_6 (a_1 ) ) $&$ E_8\times E_6 $&$ 0 $&$ 4 $\\
$ E_6 $&$ (D_4 (a_1 ) ) $&$ (E_6 (a_2 ) ) $&$ E_7\times SU(6)\times SU(2)^2
$&$
1 $&$ 2 $\\
$ E_6 $&$ (A_2^3 ) $&$ (A_5 A_1 ) $&$ E_6\times SU(3)^4 $&$ 1 $&$ 1 $\\
$ E_6 $&$ (A_2^3 ) $&$ (E_6 ) $&$ E_6\times SU(3)^4 $&$ 1 $&$ 1 $\\
$ E_6 $&$ (A_2^3 ) $&$ (E_6 (a_1 ) ) $&$ E_6\times SU(3)^4 $&$ 0 $&$ 1 $\\
$ E_6 $&$ (A_2^3 ) $&$ (E_6 (a_2 ) ) $&$ E_6\times SU(3)^4 $&$ 1 $&$ 2 $\\
$ E_6 $&$ (A_5 A_1 ) $&$ (E_6 ) $&$ E_6\times SU(2)^4\times U(1)^4 $&$ 1
$&$ 1
$\\
$ E_6 $&$ (E_6 (a_1 ) ) $&$ (A_5 A_1 ) $&$ E_6\times SU(3)^4 $&$ 1 $&$ 1
$\\
$ E_6 $&$ (A_5 A_1 ) $&$ (E_6 (a_1 ) ) $&$ E_6\times SU(3)^4 $&$ 0 $&$ 1
$\\
$ E_6 $&$ (E_6 (a_1 ) ) $&$ (E_6 ) $&$ E_6\times SU(3)^4 $&$ 1 $&$ 1 $\\
$ E_6 $&$ (E_6 ) $&$ (E_6 (a_1 ) ) $&$ E_6\times SU(3)^4 $&$ 0 $&$ 1 $\\
$ E_6 $&$ (E_6 (a_2 ) ) $&$ (E_6 ) $&$ E_6\times SU(2)^4\times U(1)^4 $&$ 1
$&$
1 $\\
$ E_6 $&$ (E_6 (a_2 ) ) $&$ (E_6 (a_1 ) ) $&$ E_7\times SO(8)\times U(1)^3
$&$
0 $&$ 1 $\\
$ E_6 $&$ (E_6 (a_1 ) ) $&$ (E_6 (a_2 ) ) $&$ E_6\times SU(3)^4 $&$ 1 $&$ 2
$\\
\hline
\end{tabular}

\end{table}


\end{document}